\documentclass[
reprint,
superscriptaddress,
nofootinbib,
nobibnotes,
 amsmath,amssymb,
 aps,
]{revtex4-1}
\usepackage{xcolor}
\usepackage{graphicx}
\usepackage{dcolumn}
\usepackage{bm}
\usepackage[bookmarksnumbered=true]{hyperref}
\hypersetup{
    colorlinks = true,
    linkcolor = violet,
    anchorcolor = violet,
    citecolor = violet,
    filecolor = violet,
    urlcolor = violet
    }
\usepackage{amsthm}
\usepackage{mathtools}
\usepackage{physics}
\usepackage{xcolor}
\usepackage{graphicx}
\usepackage[left=23mm,right=13mm,top=35mm,columnsep=15pt]{geometry} 
\usepackage{adjustbox}
\usepackage{placeins}
\usepackage[T1]{fontenc}
\usepackage{lipsum}
\usepackage{csquotes}

\usepackage{slashed}

\usepackage{tikz} 
\usepackage{tikz-feynman}
\tikzfeynmanset{compat=1.0.0}
\usetikzlibrary{shapes,arrows,positioning,automata,backgrounds,calc,er,patterns}
\usetikzlibrary{quotes,angles}
\usetikzlibrary{decorations.pathmorphing}
\tikzset{snake it/.style={decorate, decoration=snake}}

\usepackage{float}
\usepackage{capt-of}

\newcommand{\nk}{{\bf      k}}
\newcommand{\np}{{\bf      p}}
\newcommand{\nq}{{\bf      q}}

\begin{document}

\title{Towards a more complete description of nucleon distortion\\ in lepton-induced single-pion production at low-$Q^2$}

\author{J. Garc\'ia-Marcos}
\email{javier31@ucm.es}
\affiliation{Grupo de F\'isica Nuclear,\\Departamento de Estructura de la Materia, F\'isica T\'ermica y Electr\'onica,\\Facultad de Ciencias F\'isicas, Universidad Complutense de Madrid and IPARCOS\\CEI Moncloa, Madrid 28040, Spain}%
\affiliation{Department of Physics and Astronomy, Ghent University, B-9000 Gent, Belgium}%

\author{T. Franco-Munoz}
\affiliation{Grupo de F\'isica Nuclear,\\Departamento de Estructura de la Materia, F\'isica T\'ermica y Electr\'onica,\\Facultad de Ciencias F\'isicas, Universidad Complutense de Madrid and IPARCOS\\CEI Moncloa, Madrid 28040, Spain}%

\author{R. Gonz\'alez-Jim\'enez}
\affiliation{Grupo de F\'isica Nuclear,\\Departamento de Estructura de la Materia, F\'isica T\'ermica y Electr\'onica,\\Facultad de Ciencias F\'isicas, Universidad Complutense de Madrid and IPARCOS\\CEI Moncloa, Madrid 28040, Spain}%

\author{A. Nikolakopoulos}
\affiliation{Theoretical Physics Department, Fermilab, Batavia IL 60510, USA}%

\author{N. Jachowicz} 
\affiliation{Department of Physics and Astronomy, Ghent University, B-9000 Gent, Belgium}%

\author{J.M. Ud\'ias} 
\affiliation{Grupo de F\'isica Nuclear,\\Departamento de Estructura de la Materia, F\'isica T\'ermica y Electr\'onica,\\Facultad de Ciencias F\'isicas, Universidad Complutense de Madrid and IPARCOS\\CEI Moncloa, Madrid 28040, Spain}%

\date{\today} 

\begin{abstract}
Theoretical predictions for lepton-induced single-pion production (SPP) on $^{12}$C are revisited in order to assess the effect of different treatments of the current operator. On one hand we have the asymptotic approximation, which consists in replacing the particle four-vectors that enter in the operator by their asymptotic values, i.e., their values out of the nucleus. On the other hand we have the full calculation, which is a more accurate approach to the problem. 
We also compare with results in which the final nucleon is described by a relativistic plane wave, to rate the effect of the nucleon distortion.
The study is performed for several lepton kinematics, reproducing inclusive and semi-inclusive cross sections belonging to the low-$Q^2$ region (between 0.05 and 1 GeV$^2$), which is of special interest in charged-current (CC) neutrino-nucleus 1$\pi$ production. Inclusive electron results are compared with experimental data. 
We find non-trivial corrections comparable in size with the effect of the nucleon distortion, namely, corrections up to 6\%, either increasing or diminishing the asymptotic prediction, and a shift of the distributions towards higher energy transfer. 
For the semi-inclusive cross sections, we observe the correction to be prominent mainly at low values of the outgoing nucleon kinetic energy. 
Finally, for CC neutrino-induced 1$\pi^+$ production, we find a reduction at low-$Q^2$ with respect to both the plane-wave approach and the asymptotic case.
\end{abstract}

\maketitle

\section{Introduction} \label{sec:introduction}

In accelerator-based neutrino experiments, such as DUNE~\cite{DUNE16}, NO$\nu$A~\cite{NOvA20,Nova23} and MINER$\nu$A~\cite{Minerva19-1,Minerva2019-2}, inelastic interactions constitute the main interaction mechanism that contribute to the total cross sections. In other experiments such as T2K~\cite{T2K-13} or the SBN program~\cite{SBN,Micro2019,Micro2022}, quasielastic (QE) scattering is the main interaction mechanism but single-pion production (SPP) in the resonance ($\Delta$-baryon) region also plays an important role~\cite{e-vs-nu_Amaro20}. Also, SPP is a background in the QE-like or $0\pi$ signal, for example, if the pion is below the detection threshold or the resonance has a non-pionic decay. These events are modeled by event generators based on theoretical models, therefore, realistic predictions are essential to diminish systematic errors in the neutrino energy reconstruction~\cite{whitepaper2018}.
Moreover, at low-$Q^2$, model predictions systematically overshoot experimental cross sections from T2K and, mostly, MINERvA datasets~\cite{assessin_AN23}; it makes the study of model uncertainties in this region interesting. 

There are several approaches describing electroweak SPP on the nucleon~\cite{HNV1,Sato2003,Ahmad2006,Buss2007,Praet2009,Martini2009,Serot2012,Ivanov2016,Rafi2016,Minoo2018,Nakamura10}.
By kinematic constraints, the amplitudes for SPP on free nucleons can at most depend on the invariant mass $W$, the squared-four momentum transfer $Q^2$ and the scattering angles of the pion $\Omega_\pi$.
The underlying description for these amplitudes, usually in terms of nucleons, mesons and nucleon resonances, can depend explicitly on the kinematics of all external particles.
Therefore, the use of such a model for SPP in the nucleus has to deal with this dependence, and with the fact that the nucleons inside the nucleus are not fixed momentum-states, i.e., they are off-shell.
The study of this off-shellness is the main purpose of this work.
For this reason we use the model of Ref.~\cite{RGJ-HM17}. It is based on the tree-level diagrams from the non-linear sigma model Lagrangian~\cite{ChPTbook}, as used in many descriptions of SPP in the $\Delta$ region.
These diagrams make it straightforward to compute the amplitude for kinematics reached in SPP on the nucleus, and to include the off-shell features aforementioned.

The initial nucleon wave function is obtained by solving the Dirac equation with relativistic mean field (RMF) potentials~\cite{RMF}. 
For the final nucleon we work within the relativistic distorted wave impulse approximation (RDWIA)~\cite{udias93,kelly2005}, which means that the scattered nucleon wave function is also a solution of the Dirac equation in the continuum. In this work, we will use the energy-dependent RMF (ED-RMF) potential~\cite{rmf-pots}, in which orthogonality between initial- and final-nucleon states is preserved by construction. 
We describe the pion as a plane wave, 
work is in progress on implementing the distortion of the pion wave function in our framework. 

In the present work we go beyond the so-called asymptotic approximation, which is widely used, including in the Hybrid model~\cite{emspp_RGJ19,assessin_AN23}. 
It is more often called `local approximation'~\cite{Toker83,tiator84,li93,Nakamura10}, because it eliminates coordinate derivatives in coordinate space expressions. 
Analogously, in momentum space the asymptotic approximation consists in defining the hadronic operator using the asymptotic values of the particle 4-vectors (i.e., their values out of the nucleus) instead of those inside the nucleus.
In this work, all computations are performed in momentum space~\footnote{To our knowledge, the pioneering works for pion photoproduction of Refs.~\cite{eramzhyan1983,tiator84} were the first ones that computed the amplitude in momentum space.}, so the non-asymptotic (or non-local) treatment can be trivially fully implemented. 
The advantage of the asymptotic approach is that it is computationally much less demanding (this will become clear in the next section). 
In Refs.~\cite{annals81,nagl91,Toker83,tiator84,li93,Leitner09,Nakamura10} comparisons between the asymptotic approximation and the full calculation were performed for photon- and lepton-induced coherent and incoherent pion production.

In this work, for the first time, we present a study of the non-locality effects within the framework of a fully relativistic nuclear model and for incoherent single-pion electro- and neutrino-production on nuclei, in particular for $^{12}$C. 
Both inclusive and semi-inclusive differential cross sections for different lepton kinematics at low-$Q^2$ are presented. We find non-trivial differences with corrections both to the shape and strength of the cross section.

This work is organized as follows. In Sec.~\ref{sec:kin_xs} we briefly describe the SPP process: In Sec.~\ref{2a} we explain the kinematics and cross section of SPP on the nucleus; in Sec.~\ref{2b} we summarize the most important aspects about the pion model we use, and the treatment of the nuclear dynamics. Results and Conclusions are displayed in Secs.~\ref{sec:results} and~\ref{sec:conclusions}, respectively. Finally, in App.~\ref{sec:angle_trick} we provide details of a change of variables that allows for the analytic integration over one of the angles that defines the kinematics, which helps to reduce the computational effort. 
\section{Single pion production on the nucleus} \label{sec:kin_xs}

We describe the SPP process as a one-nucleon interaction instead of a many-body one, and assume that only one boson is exchanged between leptonic and hadronic systems. These two deep-rooted considerations are the so-called impulse approximation (IA) and the first-order Born approximation, respectively.

\begin{figure}[ht!]
\centering  
\includegraphics[width=0.49\textwidth,angle=0]{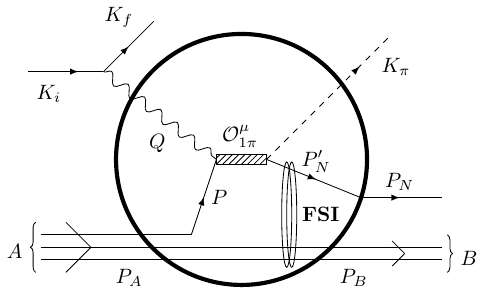}
\caption{Diagramatic representation of the general SPP process with all four-momenta depicted.}
\label{process}
\end{figure}

The process is sketched in Fig.~\ref{process}. An initial lepton with 4-vector $K_i=(E_i,\textbf{k}_i)$ goes to the final one with $K_f=(E_f,\textbf{k}_f)$ via exchange of a single boson with $Q=(\omega,\textbf{q})$. The boson couples to a bound nucleon with $P=(E,\textbf{p})$ in the nucleus $A$ with $P_A=(E_A,\textbf{p}_A)$. After the transition in which a single pion is produced, represented as $\mathcal{O}_{1\pi}^\mu$, the final state is made up of the knockout nucleon with $P_N=(E_N,\textbf{p}_N)$, the final pion with $K_\pi=(E_\pi,\textbf{k}_\pi)$ and the residual system $P_B=(E_B,\textbf{p}_B)$. For the case of the nucleon, the interaction with the residual system is taken into account, so inside the nucleus the 4-vector of the struck nucleon is $P'_N=(E_N,\textbf{p}'_N)$, being off-shell.

\subsection{Kinematics and cross section}\label{2a}

To describe the kinematics, and therefore, obtain the cross section of the SPP process, only 9 independent variables are needed~\cite{RGJ-PoS19}. We choose the following variables $(k_i,k_f,\theta_f,k_\pi,\Omega_\pi,\Omega_N,E_m)$~\footnote{The variable $E_m$ represents the missing energy, i.e., the amount of energy transferred to the residual system $B$ as internal energy.} as our 9-dimensional phase space.

The four-momentum of the exchanged boson between the lepton vertex and the hadronic vertex is given by
\begin{equation}\label{Q_mu}
    Q=K_i-K_f,
\end{equation}
its three-momentum $\textbf{q}$ is taken along the $\hat{z}$-axis, $\textbf{q}=(0,0,q)$. Imposing four-momentum conservation, we obtain, for the hadronic vertex
\begin{equation}\label{had_vertex}
    Q+P_A=K_\pi+P_N+P_B.
\end{equation}
With the initial nucleus at rest, $P_A=(m_A,\textbf{0})$, momentum and energy conservation give
\begin{equation}\label{mom_had}
    \textbf{q}=\textbf{p}_B+\textbf{p}_N+\textbf{k}_\pi,
\end{equation}
\begin{equation}\label{energy_had}
    \omega+m_A=E_B+E_N+E_\pi,\\
\end{equation}
where $E_B^2=p_B^2+m_B^2$. The mass of the residual system is related to the missing energy as $m_B=E_m+m_A-M$, being $M$ the mass of the knockout nucleon. 
From Eqs. (\ref{energy_had}) and (\ref{mom_had}) one obtains a second order equation for $\np_N$.
The explicit solution can be found in Ref.~\cite{RGJ-PoS19}, such that for certain kinematics the cross section for the two energy-momentum conserving solutions should be added incoherently.

The electroweak cross section, in the most general way~\cite{e-vs-nu_Amaro20}, is given by~\footnote{Note that nothing depends on the final lepton azimuth angle $\phi_f$.}
\begin{equation}\label{xs}
    \frac{d^{10}\sigma}{d\textbf{k}_f d\textbf{p}_N d\textbf{k}_\pi dE_m}=\frac{\rho(E_m)\mathcal{F}_{X}}{(2\pi)^8}\delta(E_N+E_\pi-\omega-E)L_{\mu\nu}H^{\mu\nu},
\end{equation}
where $E=m_A-E_B$. The function $\rho(E_m)$ represents the density of final states for the residual nucleus. The factor $\mathcal{F}_X$ is given by
\begin{equation}
    \mathcal{F}_{EM}=\frac{(4\pi\alpha)^2}{Q^4}\quad,\quad \mathcal{F}_{CC}=\frac{(G_F\cos\theta_c)^2}{2},
\end{equation}
depending on if the interaction is electromagnetic (EM) or charged current (CC), being $\alpha$ the fine-structure constant, $G_F$ the Fermi coupling constant and $\theta_c$ the Cabibbo angle. The quantity $Q^2$ is defined as positive:
\begin{equation}
    Q^2=-(K_i-K_f)^2=\textbf{q}^2-\omega^2>0.    
\end{equation}

The dimensionless lepton tensor $L_{\mu\nu}$, which depends on the type of the interaction (EM or CC), is defined as 
\begin{equation}\label{lmunu_e_w}
\begin{split}
        &L_{\mu\nu}^{EM}=\frac{1}{2E_iE_f}S_{\mu\nu}\\
        &L_{\mu\nu}^{CC}=\frac{2}{E_iE_f}(S_{\mu\nu}-ihA_{\mu\nu}), 
\end{split}
\end{equation}
where it has been separated into symmetric ($S$) and antisymmetric ($A$) tensors, given by
\begin{equation}\label{lmunu_sa}
\begin{split}
        &S_{\mu\nu}=K_{i,\mu} K_{f,\nu} + K_{i,\nu} K_{f,\mu} - g_{\mu\nu}K_i\cdot K_f,\\
        &A_{\mu\nu}=\varepsilon_{\alpha\beta\mu\nu}K_i^\alpha K_f^\beta. 
\end{split}
\end{equation}
In Eqs.~\eqref{lmunu_e_w} and~\eqref{lmunu_sa}, $g_{\mu\nu}$ is the metric tensor given by $g_{\mu\nu}=diag(+,-,-,-)$, $\varepsilon_{\alpha\beta\mu\nu}$ is the fully antisymmetric Levi-Civita tensor within the convention $\varepsilon_{0123}=+1$, $i$ represents the imaginary unit, and $h$ stands for the initial lepton helicity, $h=-1$ $(+1)$ for neutrinos (antineutrinos).

The hadronic tensor is defined for each nuclear shell $\kappa$ as
\begin{equation}\label{hmunu}
    H^{\mu\nu}_\kappa=\frac{N_\kappa}{2j+1}\sum_{m_j,s_N}(J^\mu)^\dagger J^\nu,
\end{equation}
where we are summing over all final spin states, $s_N$ being the projection of the spin of the final nucleon, and averaging over initial spin states, $m_j$ being the projection of the angular momentum $j$ of the bound state. $N_\kappa$ stands for the occupation of the nuclear shell~\footnote{
Within a pure shell model $\rho(E_m)=\sum_\kappa\delta(E_m-E^\kappa_m)$, where $E_m^\kappa$ is a fixed value for each shell, and $N_\kappa=2j+1$.}. All the nuclear information is enclosed in the hadronic current 
\begin{equation}
\label{generalcurrent}
 J^\mu = J^\mu(\kappa,m_j,s_N,Q,P_N,K_\pi).
\end{equation} 
The discussion of the hadronic current is exposed in Sec.~\ref{2b}.
Integrating the Dirac delta in Eq.~\eqref{xs} over $p_N$ we get
\begin{equation}\label{xs_int}
\begin{split}
    \frac{d^9\sigma}{dE_fd\Omega dE_\pi d\Omega_\pi\Omega_N dE_m}&=\mathcal{F}_{X}\frac{E_fk_fE_\pi k_\pi E_N p_N}{(2\pi)^8 f_{rec}}\\
    &\times \rho(E_m)\,L_{\mu\nu}H^{\mu\nu},
\end{split}
\end{equation}
where 
\begin{equation}\label{frec}
    f_{rec}=\Big|1+\frac{E_N}{E_B}\Big(1+\frac{\textbf{p}_N\cdot (\textbf{k}_\pi-\textbf{q})}{p_N^2}\Big)\Big|,
\end{equation}
is the recoil factor.

\subsection{Nuclear framework and single-pion production model}\label{2b}

The Hybrid model is included in the nuclear dynamics through the hadronic current of Eq.~\eqref{generalcurrent}, as all the nuclear information is confined in it. For the most general case, it is of the form 
\begin{equation}\label{jmu_6d}
\begin{split}
    J^\mu=&\,\frac{1}{(2\pi)^{3/2}}\int d\textbf{p} \int d\textbf{k}'_\pi \, \Bar{\psi}^{s_N}(\textbf{p}'_N,\textbf{p}_N)\phi^*(\textbf{k}'_\pi,\textbf{k}_\pi)\\
    \times&\,\mathcal{O}^\mu_{1\pi}(Q,P_N',K'_\pi)\psi_{\kappa}^{m_j}(\textbf{p})
\end{split}
\end{equation}
with $\np'_N=\nq+\np-\nk'_\pi$. Here $\mathcal{O}^\mu_{1\pi}$ represents the SPP current operator, $\psi_{\kappa}^{m_j}(\textbf{p})$ is the Fourier transform of the bound nucleon relativistic wave function in coordinate space
\begin{equation}\label{TF}
    \psi_{\kappa}^{m_j}(\textbf{p})=\frac{1}{(2\pi)^{3/2}}\int d\textbf{r}e^{-i\textbf{p}\cdot\textbf{r}}\psi_\kappa^{m_j}(\textbf{r}),
\end{equation}
$\psi_\kappa^{m_j}(\textbf{r})$ is computed within the RMF model~\cite{RMF}, which is an extension of the original Walecka $\sigma-\omega$ model~\cite{WALECKA1974}. The single particle wave function $\psi_\kappa^{m_j}(\textbf{r})$ is the solution of the Dirac equation with central and vector potentials, with well-defined energy and angular momentum.
On the other hand, $\Bar{\psi}^{s_N}(\textbf{p}'_N,\textbf{p}_N)$ is the Fourier transform of the relativistic wave function of the knockout nucleon with fixed energy and spin
\begin{equation}\label{scatt-wf}
\begin{split}
    \Bar{\psi}^{s_N}(\textbf{p}'_N,\textbf{p}_N)=&\,4\pi\sqrt{\frac{E_N+M}{2M}}\sum_{\kappa,m_j,m_l}e^{i\delta_\kappa^*}i^l\\
    \times &\langle lm_l\frac{1}{2}s_N|jm_j\rangle Y^{m_l*}_l(\Omega_{\textbf{p}_N})\psi^{m_j}_\kappa(\textbf{p}'_N),
\end{split}
\end{equation}
where $\delta_\kappa$ is the phase shift, $\langle j_1m_1j_2m_2|JM\rangle$ are Clebsch-Gordan coefficients, $Y_l^{m_l}(\Omega_{\textbf{p}_N})$ are spherical harmonics, and $\psi^{m_j}_\kappa(\textbf{p}'_N)$ is a spinor obtained as in Eq.~\eqref{TF}. For the final nucleon we use the ED-RMF potential~\cite{emspp_RGJ19}, so orthogonality between nucleon initial and final states is automatically satisfied, which is important to avoid spurious contributions to the cross section~\cite{Nikolakopoulos19,Tania23}. Finally, $\phi(\textbf{k}'_\pi,\textbf{k}_\pi)$ corresponds to the final pion wave function in momentum space. 
In the most general case, pion and nucleon in Eq.~\eqref{jmu_6d} are both off-shell. They are not pure momentum states, the momentum dependence is given by the primed momenta, while the unprimed one is the asymptotic momentum given by $p_N=\sqrt{E_N^2-M^2}$ and $k_\pi=\sqrt{E_\pi^2-m_\pi^2}$.

In this work, we describe the pion as a plane wave
\begin{equation}
    \phi(\nk'_\pi,\nk_\pi)=\sqrt{\frac{(2\pi)^3}{2E_\pi}} \, \delta^{(3)}(\nk'_\pi-\nk_\pi),
\end{equation} 
and therefore, Eq.~\eqref{jmu_6d} simplifies to
\begin{equation}\label{jmu_3d}
\begin{split}
    J^\mu=\frac{1}{\sqrt{2E_\pi}} &\, \int d\textbf{p} \, \Bar{\psi}^{s_N}(\textbf{p}'_N,\textbf{p}_N)
    \mathcal{O}^\mu_{1\pi}(Q,P_N',K_\pi)\psi_{\kappa}^{m_j}(\textbf{p})
\end{split}
\end{equation}
with $\np'_N=\nq+\np-\nk_\pi$. It is also interesting to consider the RPWIA case where both the final nucleon and pion are plane waves. By computing this, the impact of the distortion in the final nucleon can be assessed. In this scenario, the hadronic current is
\begin{equation}\label{jmu_pw}
\begin{split}
    J^\mu=\sqrt{\frac{(2\pi)^3 M}{2E_\pi E_N}}\Bar{u}(\textbf{p}_N,s_N)
    \mathcal{O}^\mu_{1\pi}(Q,P_N,K_\pi)\psi_\kappa^{m_j}(\textbf{p})
\end{split}
\end{equation}
with $\np=\np_N+\nk_\pi-\nq$. 

\,

\subsubsection{Current operator}\label{O1pi}

\begin{figure}[ht!]
\centering  
\includegraphics[width=0.24\textwidth,angle=0]{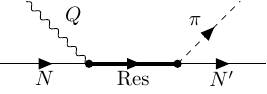}
\includegraphics[width=0.24\textwidth,angle=0]{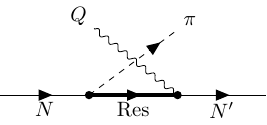}
\caption{Left: $s$-channel (resonance pole, $RP$). Right: $u$-channel (cross-resonance pole, $CRP$).}\label{res-diagrams}
\end{figure}

The current operator is constructed by summing the amplitudes coming from several Feynman diagrams. On one hand, the direct ($RP$ or resonance pole) and crossed ($CRP$ or crossed-resonance pole) diagrams for nucleon resonances can be seen in Fig.~\ref{res-diagrams}. The resonances included here are the $P_{33}(1232)$ or $\Delta$-baryon, $D_{33}(1515)$, $P_{11}(1430)$ and $S_{11}(1535)$.

\begin{figure}[ht!]
\centering  
\includegraphics[width=0.24\textwidth,angle=0]{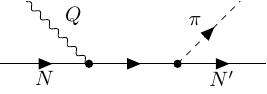}
\includegraphics[width=0.24\textwidth,angle=0]{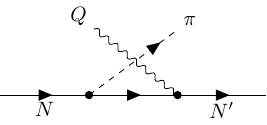}
\includegraphics[width=0.15\textwidth,angle=0]{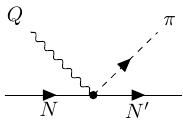}
\includegraphics[width=0.15\textwidth,angle=0]{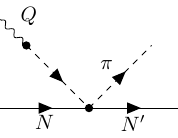}
\includegraphics[width=0.15\textwidth,angle=0]{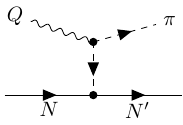}
\caption{ChPT-background diagrams (from left to right and top to bottom): $s$-channel (nucleon pole, $NP$), $u$-channel (cross-nucleon pole, $CNP$), contact term ($CT$), pion pole ($PP$), and $t$-channel (pion-in-flight term, $PF$).}\label{chpt-diagrams}
\end{figure}

Moreover, the tree level background terms derived from the $\pi N$-lagrangian of chiral perturbation theory (ChPT) are also included. The background contributions are shown in Fig.~\ref{chpt-diagrams}.

All these diagrams constitute the Hernandez, Nieves and Valverde (HNV) model of Refs.~\cite{HNV1,HNV2,HNV3}, which is valid for invariant masses $W\lesssim 1.4$ GeV, where $W=\sqrt{s}$, and $s=(P+Q)^2$. Then, this model reaches its limit of applicability as it only includes lowest-order amplitudes (see, for instance, Ref.~\cite{RGJ-HM17}). For that reason, in Ref.~\cite{RGJ-HM17}, an extension of the model based on Regge phenomenology~\cite{regge1-MV97,Regge2-Kaskulov10,Regge3-MV98,Regge4-Ryckebusch14} was presented. The Regge approach is a well-tested formalism that permits access to the high energy regime ($W>2$ GeV). The Regge phenomenology was applied by \textit{reggeizing} the ChPT-background contributions, what was denominated in Ref.~\cite{RGJ-HM17} as the ``ReChi model''. Finally, both models, HNV and ReChi, were combined by a blending function that transitions from one model to the another while $W$ increases.

The resulting Hybrid model has been used in several works, both for electro- and neutrino-production. In Ref.~\cite{RGJ-HM17}, it was tested on free nucleons. In Refs.~\cite{SPP_water-18,Modeling17,SPP-t2k-water18}, it was applied within the relativistic plane wave impulse approximation (RPWIA) to scattering on nuclei. Finally, in Refs.~\cite{emspp_RGJ19,assessin_AN23}, the distortion of the final nucleon was included.

\begin{figure}[ht!]
\centering  
\includegraphics[width=0.43\textwidth,angle=0]{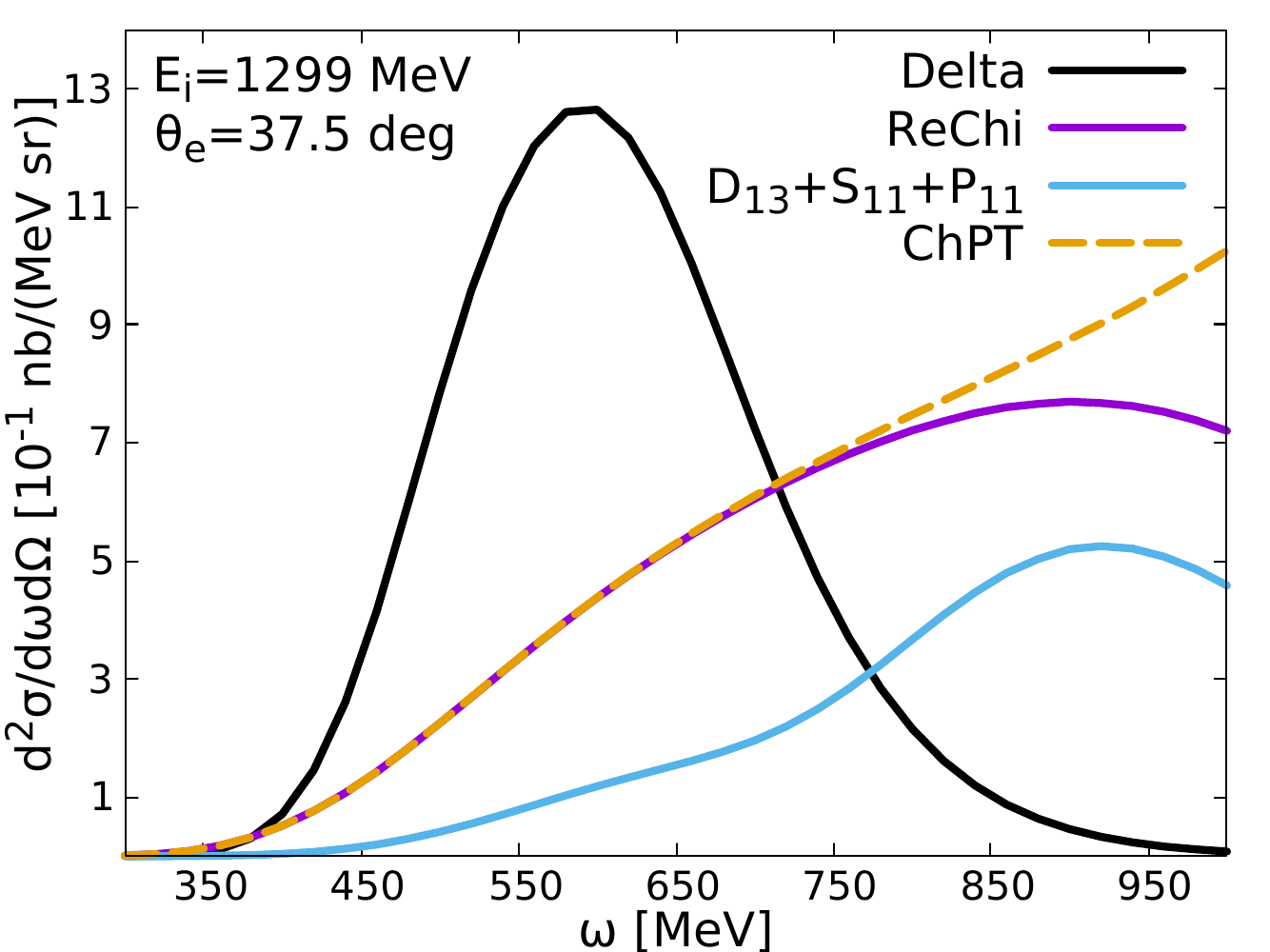}
\caption{Inclusive $^{12}$C$(e,e')$ cross section using different pieces of the current operator for a specific lepton kinematics. Black line is for the $\Delta$ contribution, purple line is for \textit{reggeized} background and blue line is the contribution from the other three resonances. Dashed orange line is the ChPT background but without Regge. Calculation performed within the RPWIA approach.}
\label{op_splitted}
\end{figure}

As an example, in Fig.~\ref{op_splitted} we show the contributions centered on the Delta (and slightly beyond it) of different parts of the operator for electron scattering. We also show the behavior of the ChPT contribution without the Regge phenomenology so one can judge its impact at high energies. 
Note that the final result is not the sum of the cross section contributions displayed separately in Fig.~\ref{op_splitted} but the coherent sum of their amplitudes, see Eq.~\eqref{op}.

The current operator of the hadronic current reads
\begin{equation}\label{op}
    \mathcal{O}^\mu_{1\pi}=\sum_R\mathcal{O}^\mu_R+\mathcal{O}^\mu_{ChPT}, 
\end{equation}
where $\mathcal{O}_R^\mu$ is the operator of resonance $R$, taking into account that $\mathcal{O}^\mu_R=\mathcal{O}^\mu_{RP}+\mathcal{O}^\mu_{CRP}$. Analogously, $\mathcal{O}^\mu_{ChPT}$ represents the sum of the different non-resonant background current operators. 

So far, all the Hybrid model predictions with nucleon distortion have been carried out using the asymptotic approximation. It consists in replacing the primed momenta in the operator by their asymptotic values
\begin{equation}\label{asymp_app}
      \mathcal{O}^\mu_{1\pi}(Q,P_N',K'_\pi)\longrightarrow \mathcal{O}^\mu_{1\pi}(Q,P_N,K_\pi),  
\end{equation}
hence, the operator does not depend on $\np$ anymore and has to be evaluated only once before the integral over $\np$ in Eq.~\eqref{jmu_3d}. 
The current operator $\mathcal{O}^\mu_{1\pi}$ is a complex object whose evaluation requires a non-negligible computational effort, so the asymptotic approximation allowed us in previous works to produce systematic comparisons with inclusive electron-nucleus and flux-folded neutrino-nucleus cross section data~\cite{emspp_RGJ19,assessin_AN23}, which otherwise would have been too computationally demanding. In the RPWIA case, where final particles are described by plane waves and therefore, they are on-shell, it is meaningless to talk about asymptotic approximation or full calculation, both are exactly equivalent.

From the explicit expressions of the diagrams in Figs.~\ref{res-diagrams} and~\ref{chpt-diagrams}, given in Ref.~\cite{RGJ-HM17}, it is easy to see that the terms that are affected by the asymptotic approximation to Eq.~\eqref{jmu_3d}, are the propagators of the direct and crossed resonances (including the nucleon pole).
For the spin-3/2 resonances, the electroweak coupling to the resonance is affected as well, through the terms proportional to $C_4^V$, $C_4^A$ and $C_5^V$.
As the pion is treated as a plane wave in this work, the contributions of the $PP$ and $PF$ terms will not change. When pion distortion would be included, only the contact terms are unaffected by the asymptotic approximation.

\begin{figure}[ht!]
\centering  
\includegraphics[width=0.43\textwidth,angle=0]{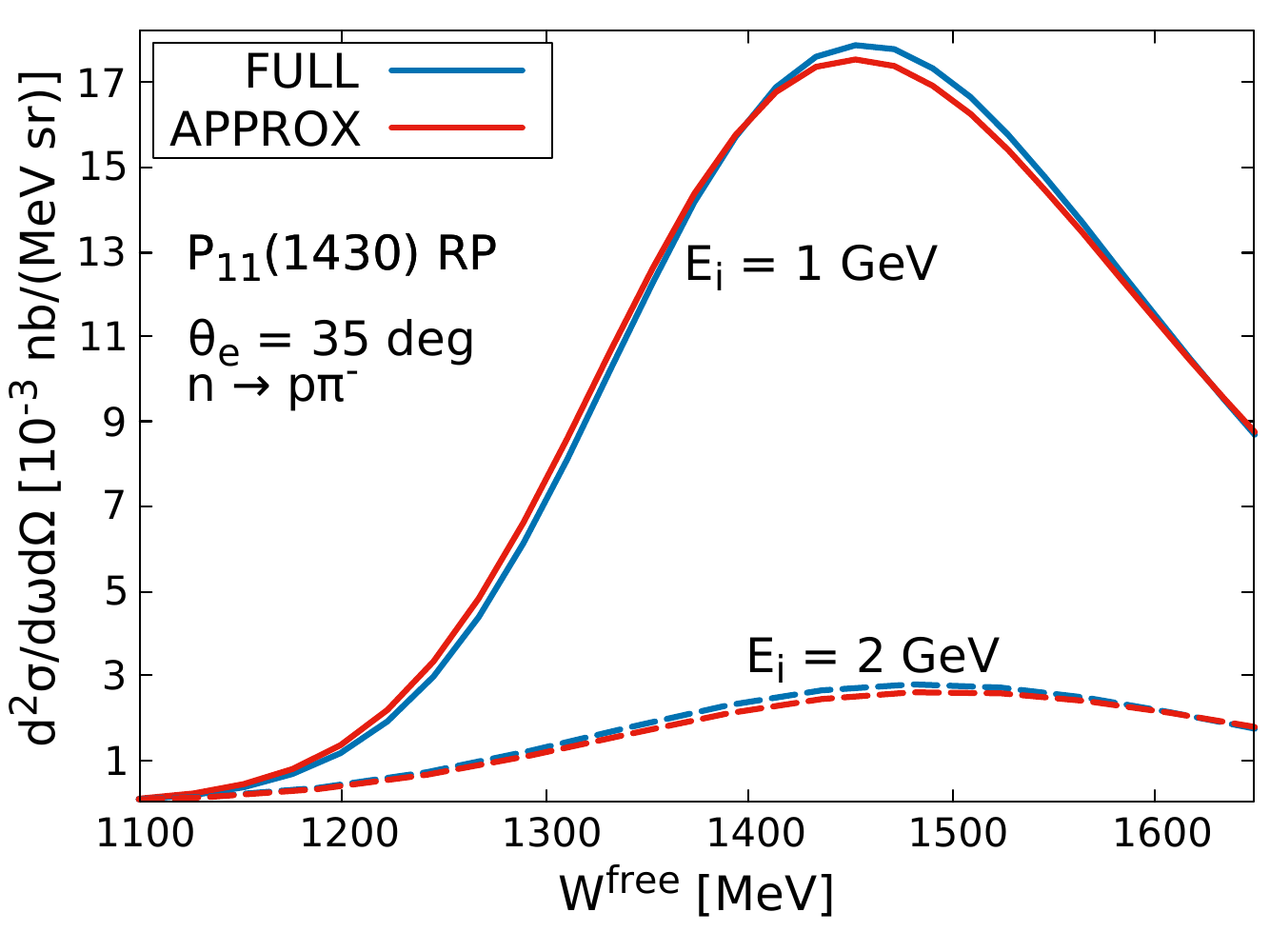}
\caption{The effect of the asymptotic approximation on the $P_{11}(1430)$ $RP$ diagram. Double differential cross sections for the reaction $^{12}$C$(e,e')$ for the channel $n\to p\pi^-$ within two different incoming energies are presented as function of $W^{free}$. Solid lines are for $E_i=1$ GeV, dashed lines are for $E_i=2$ GeV.}
\label{P11-example}
\end{figure}

As an example of how the full or approximate treatments of the operator can change each term in Eq.~\eqref{op}, in Fig.~\ref{P11-example} we show the double differential electromagnetic cross section for the $n\to p\pi^-$ channel with two different incoming energies as a function of $W^{free}=\sqrt{M_N^2+2\omega M_N-Q^2}$, i.e., the invariant mass if the interaction would occur on a free stationary nucleon. We observe that the full calculation yields to a small increment and a shift towards higher $W^{free}$ values with respect to the approximate calulation.

\section{Results} \label{sec:results}

\begin{figure*}[ht!]
\centering  
\includegraphics[width=0.49\textwidth,angle=0]{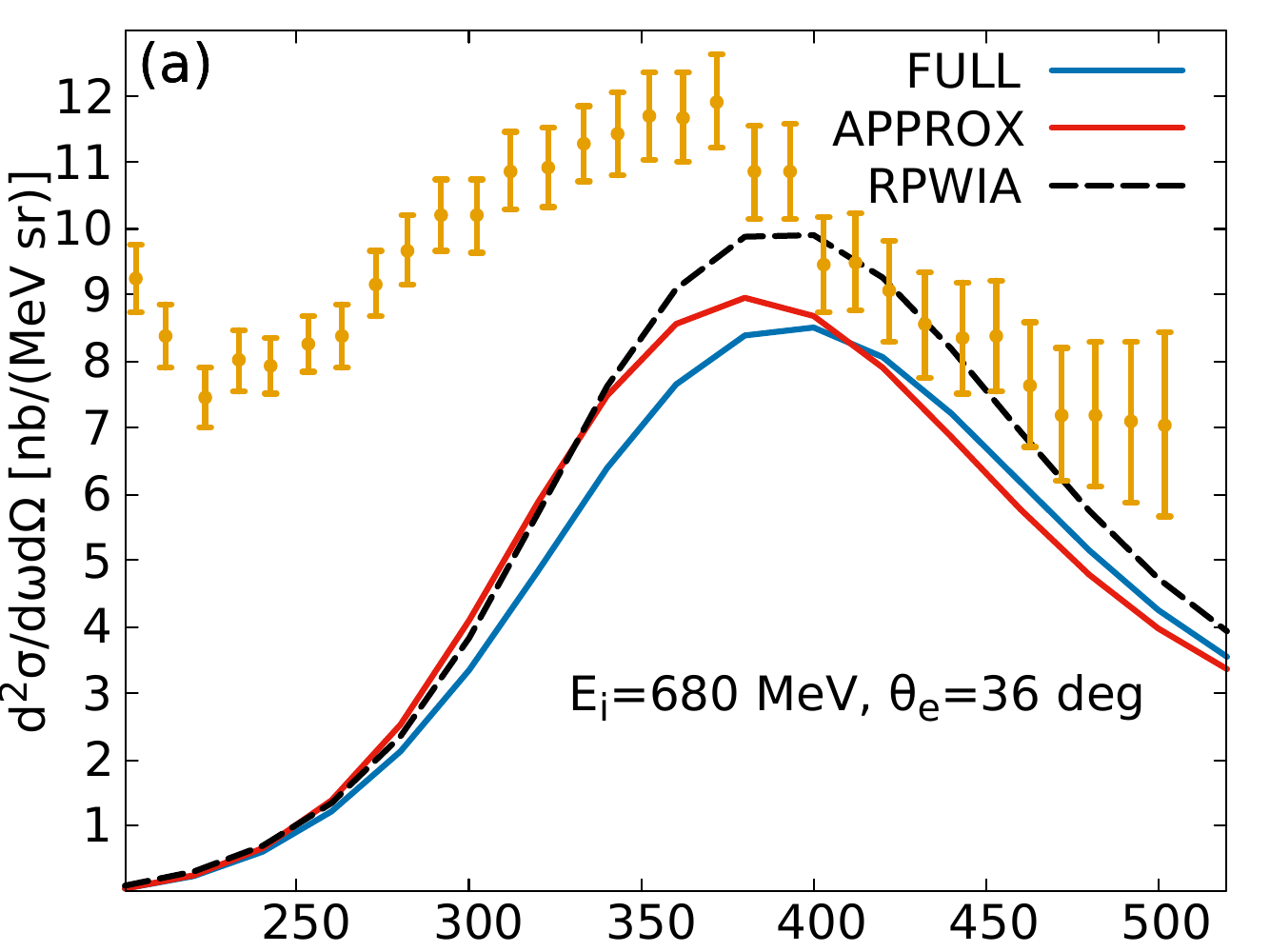}
\includegraphics[width=0.49\textwidth,angle=0]{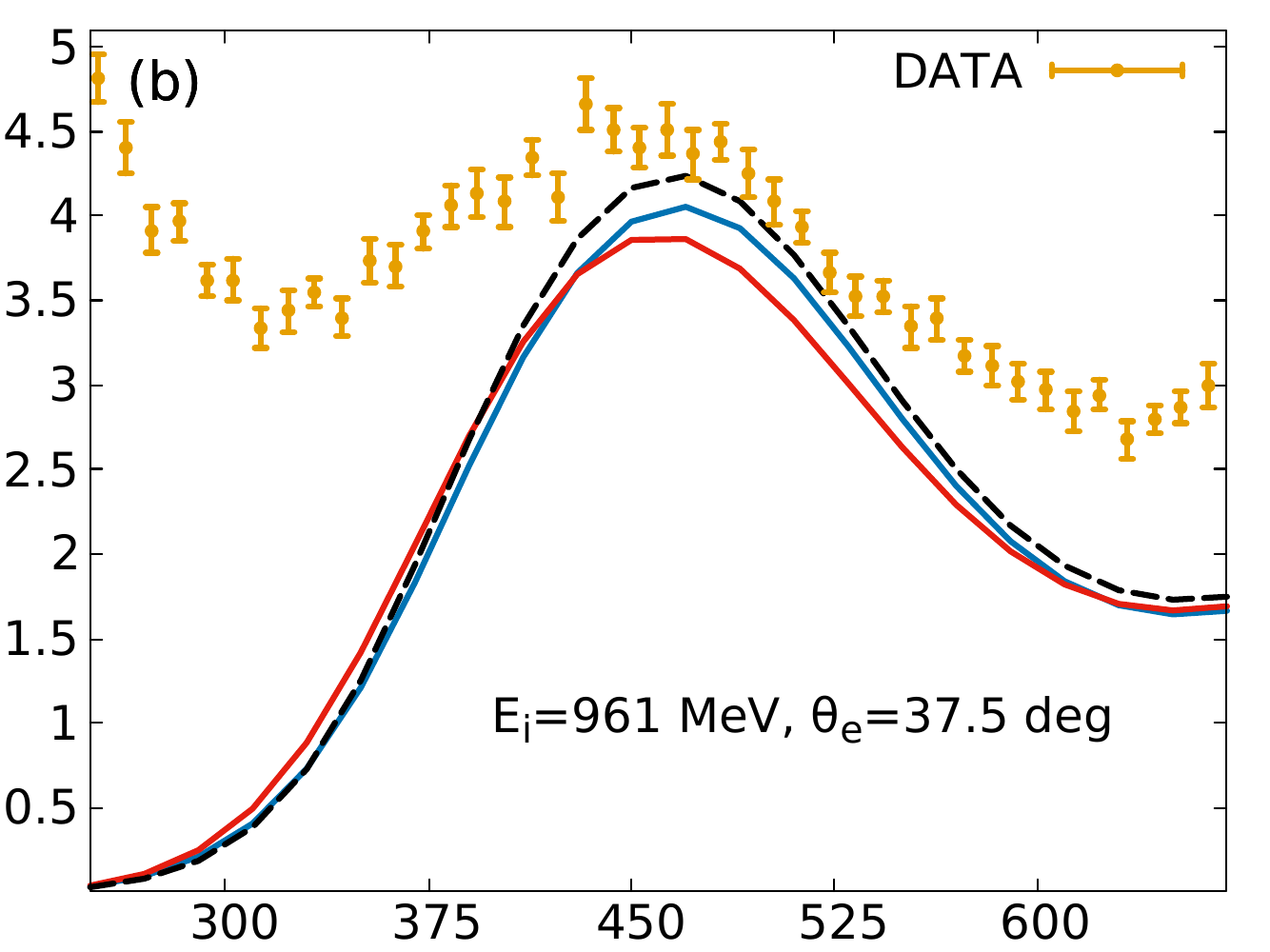}
\includegraphics[width=0.49\textwidth,angle=0]{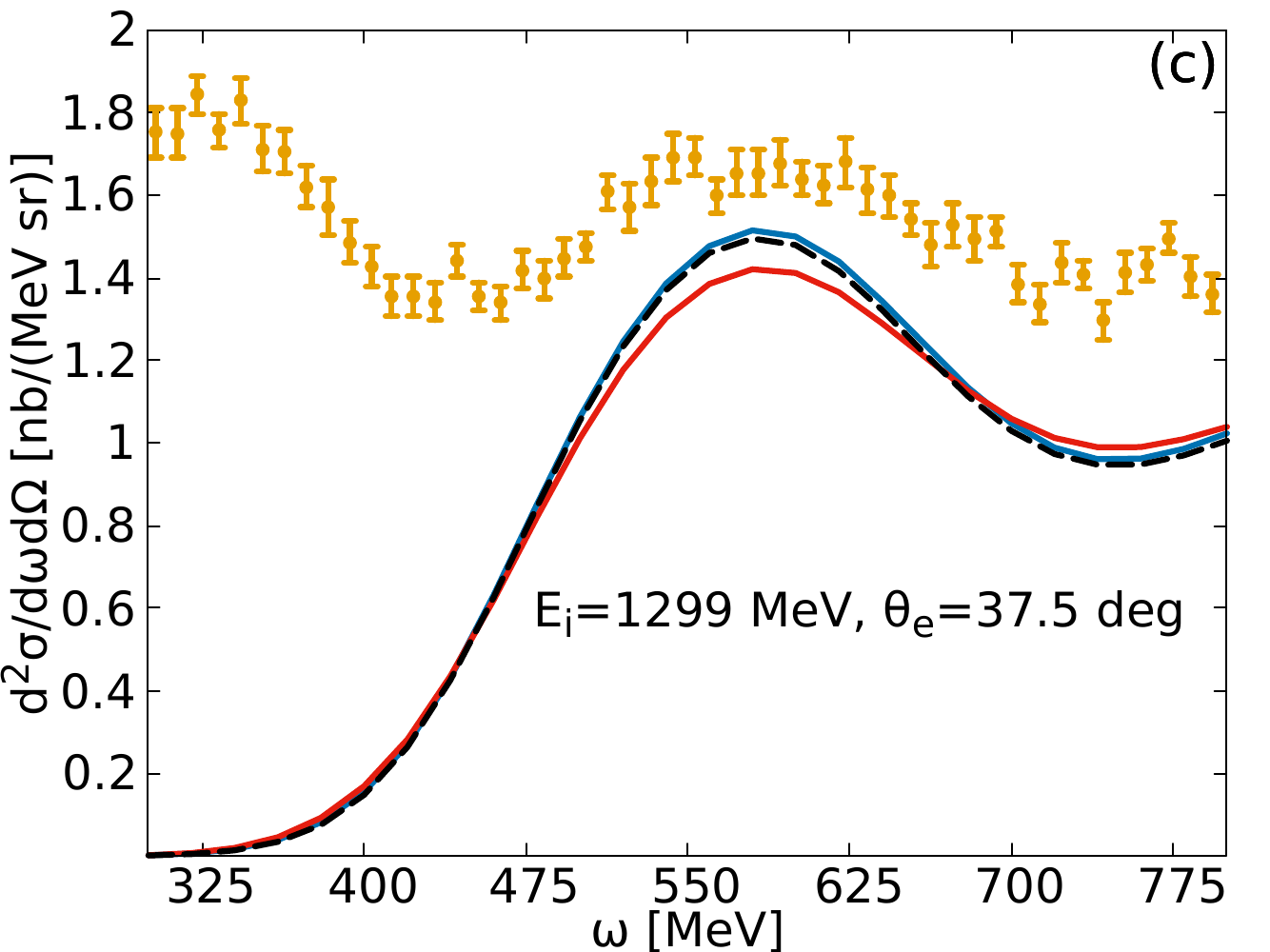}
\includegraphics[width=0.49\textwidth,angle=0]{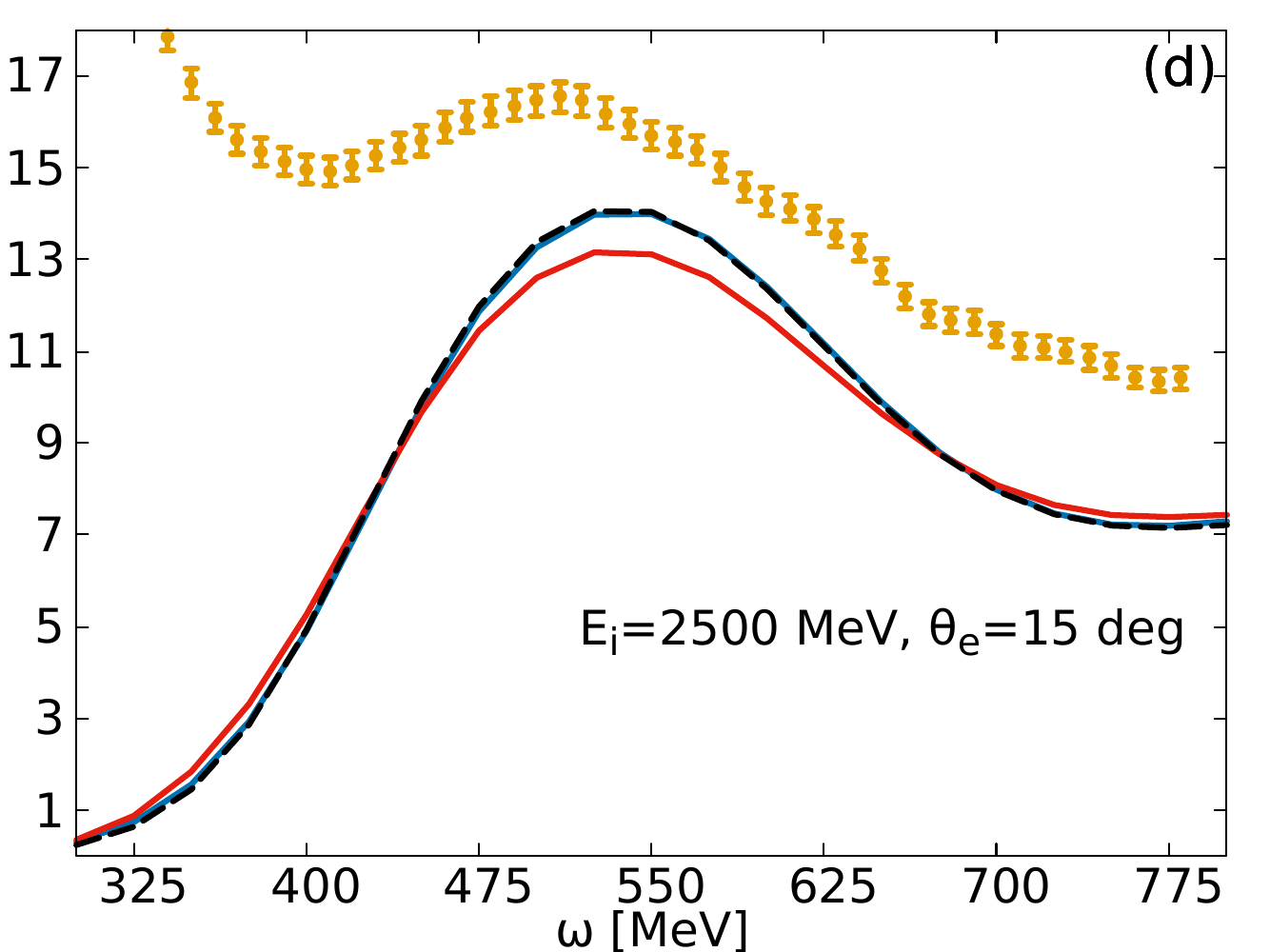}
\caption{Inclusive SPP $^{12}$C$(e,e')$ differential cross section for different kinematics. Plots show the predictions of the RPWIA, and ED-RMF models with (APPROX) and without (FULL) asymptotic approximation. Data of panel (a) are from~\cite{680data}, data of panels (b) and (c) are from~\cite{961-1299data}, and data of panel (d) are from~\cite{2500data}.
}
\label{inclusive}
\end{figure*}

Our motivation is to address the effects of the asymptotic approximation, identifying the kinematical regions where it works better and where it fails.
For that, we focus on the study of inclusive and semi-inclusive electron scattering cross sections on $^{12}$C. We also show neutrino scattering results for a fixed incoming energy. We include only the SPP channel.

\subsection{Electroproduction}\label{electro-results}

The inclusive cross section is obtained by explicit integration over the hadronic variables 
\begin{equation}\label{diff-xs}
    \frac{d^3\sigma}{d\omega d\Omega}=\int dT_N d\Omega_N d\Omega_\pi \frac{d^8\sigma}{d\omega d\Omega dT_N d\Omega_\pi d\Omega_N }.
\end{equation}

In Fig.~\ref{inclusive} we show our predictions for the inclusive cross section and compare them with experimental data. We present RDWIA with and without the asymptotic approximation, and RPWIA. For the nucleon distortion, we have considered the ED-RMF potential; though not shown here, we also performed calculations with the real part of the energy dependent $A$-independent carbon 12 potential (EDAI-C) of Ref.~\cite{cooper93}, and found only slight differences with respect to ED-RMF, mainly in the low $T_N$ region, as expected~\cite{rmf-pots,Franco-Munoz22}.

The panels show three different kinematics. First, we point out that an underprediction of the experimental data is expected, as other reaction channels contributing to the experimental signal, like quasielastic scattering, multinucleon knockout, two-pion production, among others, are not included. 

In Ref.~\cite{emspp_RGJ19} it was found that, within the asymptotic approximation, the distortion of the final nucleon resulted in a reduction of the total strength and a shift of the distributions towards lower $\omega$ values, with respect to the RPWIA predictions that is taken as reference.
Here, we find that with the full calculation (i.e., RDWIA and without asymptotic approximation) the reduction of the strength tends to remain but the shift disappears.

At high energy and momentum transfer, which corresponds to high kinetic energy of the knocked out nucleon, the three approaches must tend to move closer to each other, because the energy dependent potentials weaken for increasing nucleon energies~\cite{emspp_RGJ19}.
This is confirmed by the results in Fig.~\ref{inclusive}, where we observe that the predictions from the three models are quite different at low energies, panel (a), but they tend to get closer for higher energies, panel (c) and (d). 

It is interesting to observe that, for the kinematics of Fig.~\ref{inclusive}(c) and (d), the full model is extremely close to the much simpler RPWIA one. It would be dangerous to understand from this that the RPWIA treatment is compatible with the more complete RDWIA approach.  
From the results in Fig.~\ref{inclusive}, we do conclude that the impact of the approximations in the $1\pi$ production operator are comparable to these of nucleon distortion and around or below 10\% for these inclusive results, depending on the kinematics.

\begin{figure*}[ht!]
\includegraphics[width=1\textwidth,angle=0]{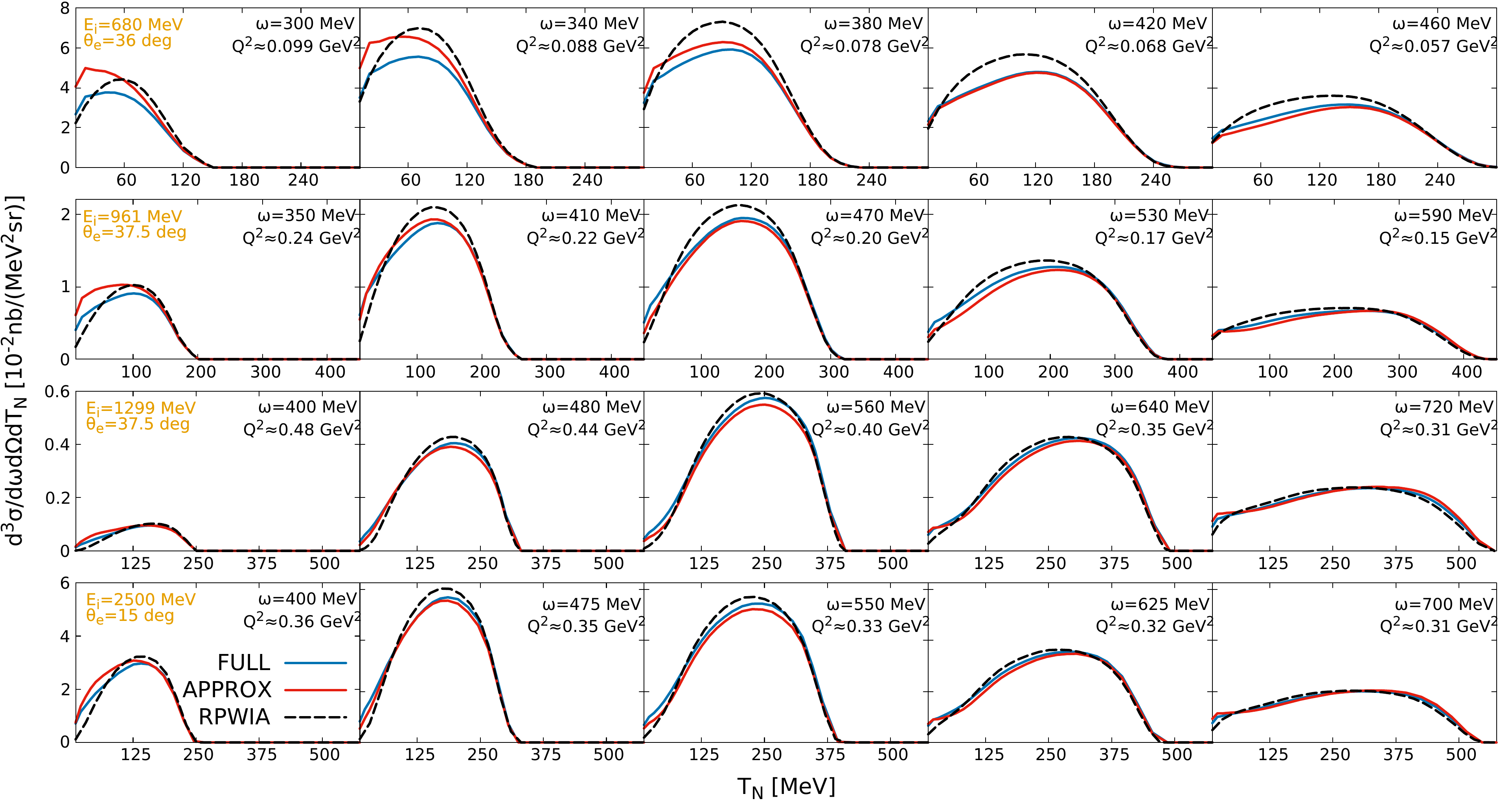}
\caption{Semi-inclusive SPP $^{12}$C$(e,e')$ differential cross section for different kinematics. Plots show RPWIA, and ED-RMF (with and without asymptotic approximation) treatments for the final nucleon. Each row is for a different kinematic, and the energy transfer $\omega$ is fixed for each pannel.}
\label{TN_multi}
\end{figure*}

To better understand the effect of the full calculation, we present in Fig.~\ref{TN_multi} semi-inclusive differential cross sections as a function of the kinetic energy of the nucleon $T_N$. 
The semi-inclusive cross section is obtained integrating over the pion and nucleon solid angles. 
We find $T_N$ to be the most relevant variable as the nuclear potential felt by the final nucleon depends on it. First, second, third and fourth rows correspond to the lepton kinematics of Fig.~\ref{inclusive} (a), (b), (c) and (d), respectively.
In every row, we show the results from low to high $\omega$ values in regular steps, and we set the same scale for $x$- and $y$-axes to assess the actual strength that contributes to the inclusive cross section. 
We also give the $Q^2$ value in each case. 

We find that, in general, the full result and the one with asymptotic approximation have similar shapes, determined by the distortion of the nucleon. 
We find that the full calculation is always lower than the approximate one up to $\omega\approx 430$ MeV, then it is always larger up to $\omega\approx 655$ MeV, where the relative magnitude switches again.
This can explain, in part, that for low incident energy we have a reduction of the inclusive cross section and as the incident energy increases, the situation is reversed. 

The three models tend to overlap as $T_N$ grows, where the distortion effect diminishes. 
At low $T_N$ ($<100$ MeV) we find large differences between the three approaches; this is relevant when the cross section is large in that $T_N$ region, as is the case of first row in Fig.~\ref{TN_multi}, which corresponds to Fig.~\ref{inclusive}(a), but irrelevant when the cross section is small for those $T_N$ values. 

\subsection{Neutrino CC1$\pi^+$ production}

\begin{figure*}[ht!]
\centering 
\includegraphics[width=0.49\textwidth,angle=0]{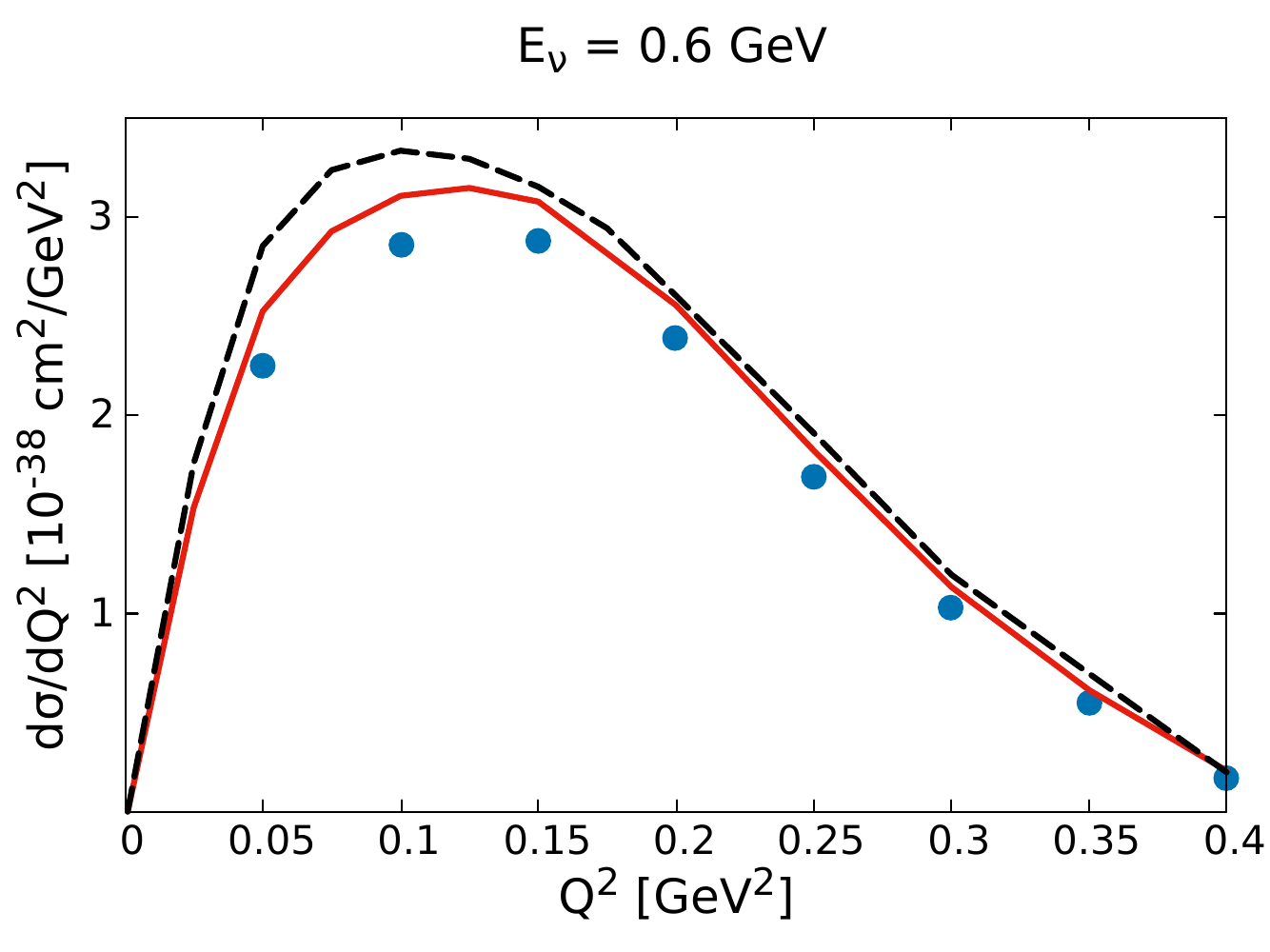}
\includegraphics[width=0.49\textwidth,angle=0]{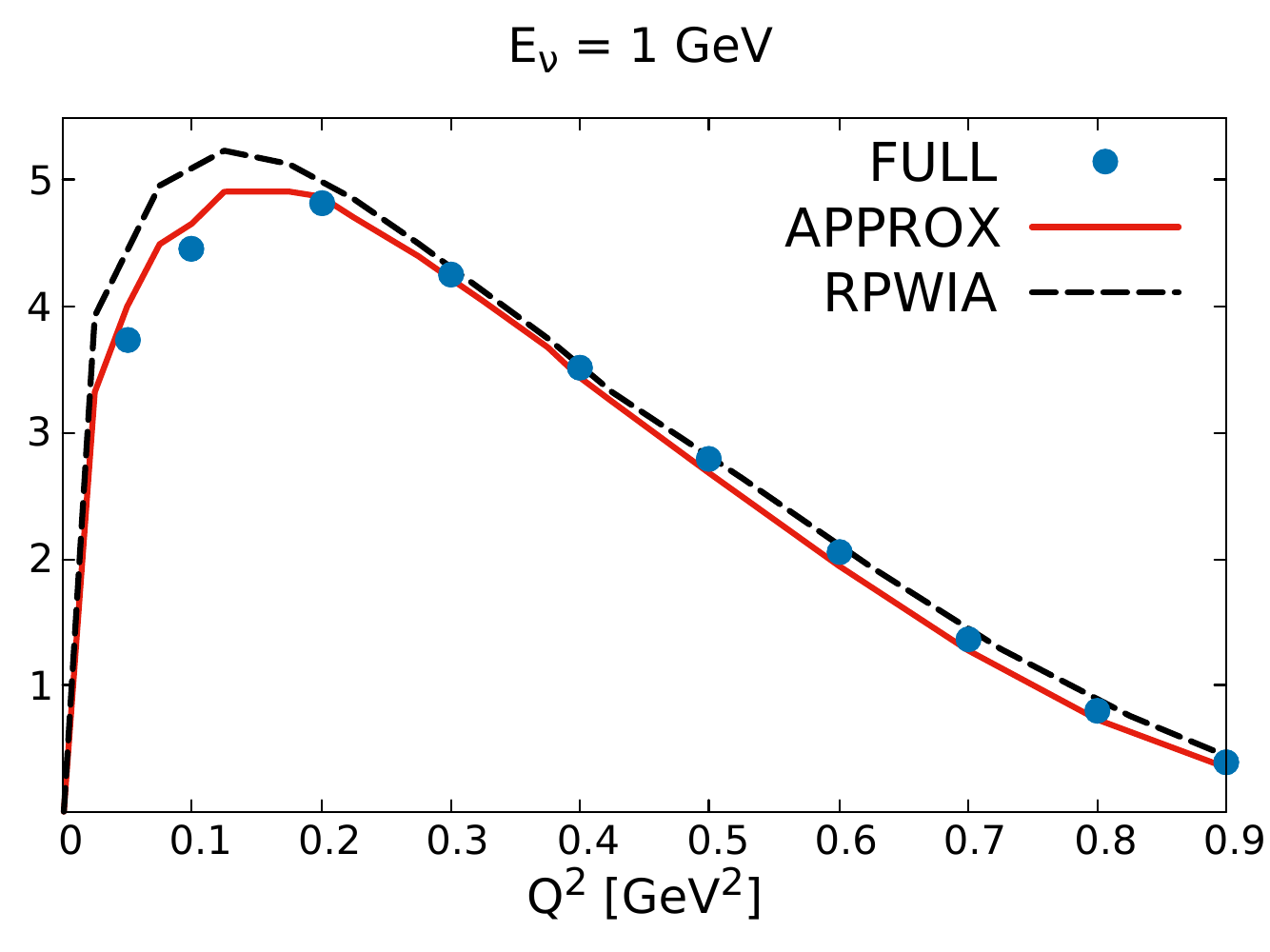}
\caption{Single $\pi^+$ production off $^{12}$C in terms of $Q^2$ for two neutrino energies: $E_\nu=0.6$ GeV and $E_\nu=1$ GeV. Blue points represent ED-RMF without asymptotic approximation and red lines represent ED-RMF with asymptotic approximation. Dashed black lines stand for RPWIA.}
\label{Q2-results}
\end{figure*}

We have computed CC $\nu_\mu$-induced 1$\pi^+$-production on $^{12}$C differential cross section as a function of $Q^2$ to assess the effect of the full calculation at the low-$Q^2$ region in the neutrino sector, where also the axial part of the current operator contributes. 
The double differential cross section as a function of $\omega$ and $Q^2$ reads
\begin{equation}\label{dQ2-jac}
    \frac{d^2\sigma}{d\omega dQ^2}=\frac{\pi}{E_\nu k_\mu}\times\frac{d^2\sigma}{d\omega d\cos\theta_\mu}.
\end{equation}
The squared four-momentum transfer is given by $$Q^2=2E_\nu(E_\mu-k_\mu\cos\theta_\mu)-m_\mu^2.$$ 
The final single differential cross section is obtained by explicit integration over $\omega$ in Eq.~\eqref{dQ2-jac}. Apart from full or approximated calculations within the RDWIA approach, we also show the RPWIA to account for the effect of the nucleon distortion, as in the $^{12}$C$(e,e')$ results.

In Fig.~\ref{Q2-results} we show the single differential cross section as function of $Q^2$ for different incoming neutrino energies. We find a reduction in the low-$Q^2$ region. However, this reduction gets smaller as the incoming energy increases, because the kinetic energy of the final nucleon $T_N$ is less restricted to low values. In general, we obtain a slight shift towards higher $Q^2$ values. The neutrino $Q^2$-distribution is a topic that raises a lot of interest in the neutrino community. The MINER$\nu$A collaboration reported a strong deficit of pion production at low-$Q^2$, where a suppression is implemented \textit{ad hoc} in that region in order to get agreement with the data~\cite{Minerva-lowQ2}. Note that within our full calculation we have a reduction specifically in this region, so the effect of the nucleon distortion is incremented with respect to the plane-wave approach. We expect the pion distortion to reduce the strength of the cross section even more, in particular, at low $T_\pi$, which (ignoring nuclear recoil) corresponds to high $T_N$ .  
\section{Conclusions} \label{sec:conclusions}

In this work, we have evaluated the impact of using an approximated treatment of the hadronic current for SPP, in the context of electron and neutrino scattering off $^{12}$C. In particular, we have compared the results obtained with a local (or asymptotic approximation) and a non-local current operator. This study is of relevance because the asymptotic approximation, which makes calculations computationally more tractable in distorted-wave approaches, has been used in the past, by our group and others. 

For the electromagnetic interaction, we show results for four different lepton kinematics comparing the $\text{RPWIA}$ and the ED-RMF with and without the asymptotic approximation in the SPP operator. Non-trivial differences are found between the three approaches. 
The most prominent features are that the two RDWIA approaches provide a reduction of the strength with respect to the plane-wave picture at low and moderate energy-momentum transfer; and that the position of the peak of the cross section for the full model agrees well with the peak position from RPWIA, while with the asymptotic approximation one observes a shift towards lower $\omega$ values.

For increasing incident energy (and hence inscreasing energy-momentum transfer), the three models tends to get closer to each other, as expected. 

While results from only the inclusive cross section seem to imply that the full calculation is closer to the RPWIA than the approximate results, this is not the case for the semi-inclusive cross section. In this case we see that the approximated and full RDWIA models, in fact, are close to each other particularly in shape, with RPWIA the most different one. The difference at low $T_N$ is most apparent.

In the neutrino sector, where for the first time this effect has been studied on $^{12}$C, we find corrections to the differential cross section similar to the electroproduction case. We see a reduction at low $Q^2$ and a mild shift towards higher $Q^2$ values compared to the asymptotic approximation. These changes are more noticeable as the incoming energy decreases. This implies a larger difference between RDWIA and RPWIA treatments for the final nucleon, which shows the importance of taking into account nuclear effects and FSI. 

Overall, we find the impact of this effect to be important to describe lepton-induced SPP cross section data, either inclusive or semi-inclusive. 
In particular, this effect is more prominent at low energies.

The distortion and Pauli exclusion principle can only be correctly addressed in a fully quantum mechanical framework, we find that these nuclear effects play an important role in the interpretation of neutrino-nucleus interactions, specially at low and moderate energy and momentum transfer or, equivalently, at low-$Q^2$. 

The next step is to develop the RDWIA formalism for the final pion, and test the effect together with the other ingredients of the nuclear matrix elements.

\,

\label{sec:acknowledgements}
\begin{center}
    \textbf{Acknowledgements}
\end{center}
This work was supported by the Madrid Government under the Multiannual Agreement with Complutense University in the line Program to Stimulate Research for Young Doctors in the context of the V PRICIT (Regional Programme of Research and Technological Innovation), project PR65/19-22430; by project PID2021-127098NA-I00 funded by MCIN/AEI/10.13039/501100011033/FEDER,UE; by project RTI2018-098868-B-I00 (MCIN/AEI,FEDER,EU), and by the Fund for Scientific Research Flanders (FWO). The computational resources (Stevin Supercomputer Infrastructure) and services used in this work were provided by the VSC (Flemish Supercomputer Center), funded by Ghent University, FWO and the Flemish Government; and Brigit, the HPC of the Complutense University of Madrid.

\,

\appendix
\section{Analytical integration of one azimuth angle} \label{sec:angle_trick}
We provide the explicit expressions of a simple change of variable which allows to integrate one phase space azimuth angle analytically. This is a very useful tool, especially when the phase space is vast, as in the pion production regime. 
In the reference frame where $\textbf{q}=(0,0,q)$ (denoted as $\{\hat{x},\hat{y},\hat{z}\}$), the three-momenta of the final nucleon and pion read
\begin{equation}\label{3mom_xyz}
    \begin{split}
        \textbf{p}_N^{(xyz)}=&\,p_N(\sin\theta_N \cos\phi_N,\,\sin\theta_N\sin\phi_N,\,\cos\theta_N),\\
        \textbf{k}_\pi^{(xyz)}=&\,k_\pi(\sin\theta_\pi \cos\phi_\pi,\,\sin\theta_\pi\sin\phi_\pi,\,\cos\theta_\pi).\\
    \end{split}
\end{equation}
The auspicius variable transformation \cite{phi-Donnelly85,Donnelly85b} will be
\begin{equation}\label{change}
    \phi=\frac{\phi_\pi+\phi_N}{2}\qquad,\qquad\Delta\phi=\phi_N-\phi_\pi,
\end{equation}
where $\phi\in (0,2\pi]$ and $\Delta\phi\in (-2\pi,2\pi]$.

The inverse transformation is therefore
\begin{equation}\label{new_angles}
    \phi_\pi=\phi-\frac{\Delta\phi}{2}\qquad,\qquad\phi_N=\phi+\frac{\Delta\phi}{2}.
\end{equation}
\begin{figure}[ht!]
\centering  
\includegraphics[width=0.49\textwidth,angle=0]{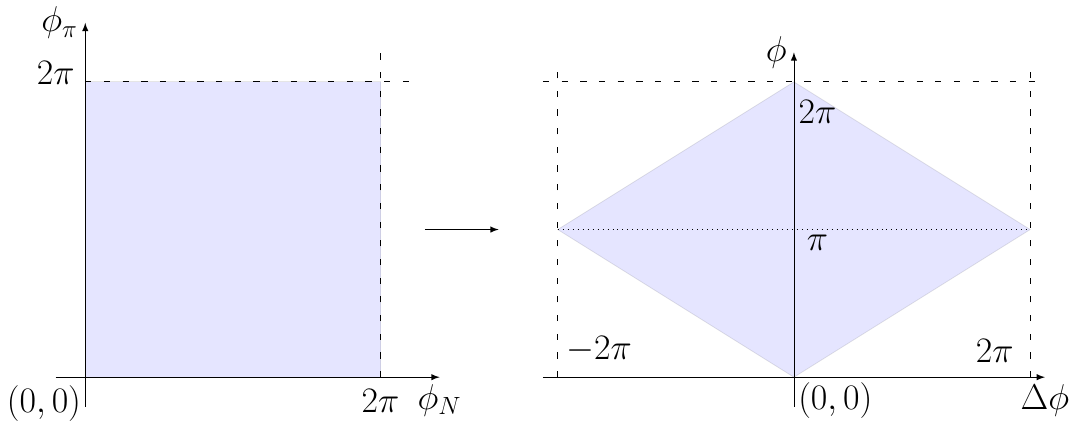}
\caption{Sketch of the domain (shaded area) of the variables before and after the transformation.}
\label{limits}
\end{figure}   

From Eqs.~\eqref{new_angles} and~\eqref{3mom_xyz}, we obtain
\begin{widetext}
\begin{equation}\label{3mom_xyz_new}
     \begin{split}
        \textbf{p}_N^{(xyz)}=&\,p_N\Big(\sin\theta_N (\cos\phi\cos\frac{\Delta\phi}{2}-\sin\phi\sin\frac{\Delta\phi}{2}),\,\sin\theta_N(\sin\phi\cos\frac{\Delta\phi}{2}+\cos\phi\sin\frac{\Delta\phi}{2}),\,\cos\theta_N\Big),\\
        \textbf{k}_\pi^{(xyz)}=&\,k_\pi\Big(\sin\theta_\pi (\cos\phi\cos\frac{\Delta\phi}{2}+\sin\phi\sin\frac{\Delta\phi}{2}),\,\sin\theta_\pi(\sin\phi\cos\frac{\Delta\phi}{2}-\cos\phi\sin\frac{\Delta\phi}{2}),\,\cos\theta_\pi\Big).\\
    \end{split}   
\end{equation}
\end{widetext}
Rotating the whole hadronic system an angle $\phi$ along $\hat{z}$ (this reference system is denoted as $\{\hat{1},\hat{2},\hat{3}\}$) we get for the three-momenta of the final hadrons
\begin{equation}\label{3mom_123_new}
     \begin{split}
        \textbf{p}_N^{(123)}=&\,p_N(\sin\theta_N\cos\frac{\Delta\phi}{2},\,\sin\theta_N\sin\frac{\Delta\phi}{2},\,\cos\theta_N),\\
        \textbf{k}_\pi^{(123)}=&\,k_\pi(\sin\theta_\pi\cos\frac{\Delta\phi}{2},\,-\sin\theta_\pi\sin\frac{\Delta\phi}{2},\,\cos\theta_\pi),\\
    \end{split}   
\end{equation}
where none of both depend on $\phi$. Finally, the hadronic current in the original reference frame, expressed in terms of the current in the new one, reads
\begin{equation}\label{newcurrent}
\begin{aligned}
            J_0&=J_0',\\
            J_1&=\cos\phi \,J_1'-\sin\phi \,J_2',\\
            J_2&=\sin\phi \,J_1'+\cos\phi \,J_2',\\
            J_3&=J_3',
\end{aligned}
\end{equation}
being $J_\mu\equiv J_\mu^{(xyz)}$ and $J_\mu'\equiv J_\mu^{(123)}$.  Following Eq.~\eqref{newcurrent}, it is straightforward to obtain the hadron tensor in $\{\hat{x},\hat{y},\hat{z}\}$ as a linear combination of the hadron tensor in $\{\hat{1},\hat{2},\hat{3}\}$.
The dependence on $\phi$ has factorized and then can be integrated analytically. For the analytic integration over $\phi$ one must take into account that the Jacobian for the transformation $\{\phi_\pi\in(0,2\pi],\phi_N\in(0,2\pi]\}$ to $\{\phi\in(0,2\pi],\Delta\phi\in(-2\pi,2\pi]\}$ is 1 and the integration limits for an integral over $\phi$ depend on $\Delta\phi$, as is sketched in Fig.~\ref{limits}.

This has been for the particular case of two azimuth angles as we have two particles in the final state. However, this can be trivially extended to an $N$-particle final state, with $N$ azimuth angles $\phi_1,...,\phi_N$, being always possible to integrate one of them analytically. Analogously to Eq.~\eqref{change}, we will have
\begin{equation}\label{nchange}
    \phi=\frac{1}{N}\sum_{i=1}^N\phi_i\quad ,\quad
    \Delta\phi_{1i}=\phi_1-\phi_i ,
\end{equation}
for $i=2,...,N$.
Thus, the new $N$ variables are $\phi,\,\Delta\phi_{12},...,\,\Delta\phi_{1N}.$ Now, the procedure is the same as in the two angles case.

\newpage
\bibliography{bibliography}

\end{document}